\documentstyle[prl,aps,psfig,multicol]{revtex}
\begin{document}
\draft

\title{Maximum Metallic Conductivity in Si-MOS Structures}
\author{V.~M.~Pudalov$^{(a,b)}$, G.~Brunthaler$^{(b)}$,
A.~Prinz$^{(b)}$, G.~Bauer$^{(b)}$}
\address{$^{(a)} $P.\ N.\ Lebedev Physical Institute of the
Russian Academy of Sciences, Moscow, Leninsky prosp. 53, Russia \\
$^{(b)}$ Institut f\"{u}r Halbleiterphysik,
Johannes Kepler Universit\"{a}t,
Linz, A-4040, Austria}
\date{\today}
\maketitle

\begin{abstract}
We found that  the  conductivity of the
two-dimensional electron system in Si-MOS structures
is limited to a maximum value, $G_{\rm max}$,
as either density increases or  temperature decreases.
This value $G_{\rm max}$ is weakly disorder dependent
and ranging from 100 to $140 e^2/h$ for samples
whose mobilities differ by a factor of 4.
\end{abstract}
\begin{multicols}{2}

According to the conventional theory of metals \cite{abra79},
the conductivity of the two-dimensional carrier system
should vanish in the limit of zero temperatures.
Recently, an unconventional metallic-like temperature
dependence of the conductivity was found
in two-dimensional (2D) carrier systems in different materials
\cite{krav94,experiments}.
The effect manifests itself in the
exponentially strong rise of the conductivity $G$
(by about one order of
magnitude in Si-MOS structures where it
is most pronounced \cite{krav94})
as the temperature decreases below
$\sim 0.3E_{F}/k_{B}$ \cite{pudal97a}.
The origin of the effect remains under
discussion \cite{theory} and is intimately related
to a question on the  ground state conductivity in the $T=0$ limit.
The existing experiments are taken at finite
temperatures (though much less than $E_F/k_{B}$) and
it is not absolutely clear whether or not the observed
``metallic-like'' temperature behavior of $G$ corresponds
to the ground state conductivity.
Since for the Fermi liquid the only possibility is
$G=0$, it was suggested,  that
the two-dimensional strongly interacting carrier system
can become a perfect metal with infinite conductivity $G$, at $T =0$
\cite{abrahams} but exhibiting non-Fermi-liquid behavior.
It was even suggested  that the 2D interacting system 
could become a superconductor \cite{super}.

In order to verify  these possibilities,
we have extended the measurements to carrier 
densities about $100$ times higher than the critical conductivity
$n_{\rm c}$, at which the exponential decrease of the resistivity sets 
in \cite{krav94,experiments}. 
Our investigations are motivated by the fact that as density increases,
the Drude conductivity  increases and ``disorder''
($1/k_F l $) decreases. From the measurements at high density,
we expected to verify whether or not
the metallic like conductivity survives
at high $G$ values, to probe the role of
Coulomb interaction effects (where the ratio of the
Coulomb to Fermi energy decreases proportionally to $n^{-1/2}$)
and of spin-related effects (which should persist
as density increases).

We have found that the conductivity in (100) Si-MOS structures
shows a \emph{maximum} as a function of carrier density.
The maximum value, $G_{\rm max} \approx 100-140$ is weakly dependent on
the mobility of the sample (conductivity throughout
this paper is in units of
$e^2/h=1/25813$\,Ohm$^{-1}$,
and the resistivity $\rho =1/G$).
The strong exponential dependence of $G(T)$ (with $dG/dT <0$)
which exists at relatively high temperatures $ T \leq 0.3E_F/k_{B}$
was found to persist up to the highest density studied. However,
at low temperatures, $T < 0.007E_F/k_{B}$ (and at least for high densities),
in the vicinity of $n=n_{\rm max}$, this
metallic like dependence transforms
into a  weak $\ln T $ dependence with a positive derivative,
$dG/dT >0$, thus indicating
the onset of a weakly localized state.

The ac- and dc-measurements
of the conductivity  were performed on
(100)~Si-MOS structures at
low dissipated power. Five samples  were studied
in the density range 0.8 to
$100 \times 10^{11}$\,cm$^{-2}$; their relevant parameters
are listed in Table~1.
In order to adjust the biasing current
such as not to destroy the  phase coherence
in the carrier system, we determined 
the phase breaking time, $\tau_{\phi}$,
from the weak negative magnetoresistance \cite{altshuler}
in low magnetic fields.
Measurements were taken in the temperature range 0.29
to 45\,K, and, partly, 0.018 to 4\,K, by sweeping slowly
the temperature during several hours. The data taken on
all five samples were qualitatively similar.

A typical density dependence of the conductivity in
the ``metallic'' range, $n= (6 -100) \times 10^{11}$\,cm$^{-2}$,
is shown in Fig.~1 for different temperatures, 0.3 to 41\,K.
The conductivity, $G $, first  increases with density,
reaches a maximum at $n=(35 - 43) \times 10^{11}$\,cm$^{-2}$,
and then decreases again. Shubnikov-de Haas data taken on
a few high mobility samples show the onset of a second frequency at
$n \geq 55 \times 10^{11}$\,cm$^{-2}$,
which is due to population of the second subband.
The reversal of the density
dependence of the conductivity  may be caused
by an increase of the scattering rate as $E_F$ approaches
the bottom of the next subband. 
Table I shows that the maximum conductivity value
is weakly dependent on disorder,
$G= 100 - 140$ for the studied samples.
At the same time, the density values $n_{\rm max}$,
corresponding to the maximum conductivity, increase
by a factor 2 as the mobility decreases by a factor 4.

In Fig.~2, the temperature dependence of the conductivity
is shown for high densities, $(8 - 80)\times 10^{11}$\,cm$^{-2}$.
As density increases, the conductivity, first increases
(the curves {\em 1} to {\em 6}),
reaches a maximum (the curve {\it 6}) at a density $n_{\rm max}$
(which is  $32 \times 10^{11}$\,cm$^{-2}$  for Si-15a),
and, finally, decreases with density
(curves {\it 7 - 12}).
This leads to a crossing of the $G(T)$-
curves taken at different
densities $n > n_{\rm max}$.
Such a crossing has also been  reported to occur for 
$p$-GaAs/AlGaAs in Ref. \cite{hamilton}. 
However, in our measurements,
the $G(n)$ curves for different temperatures,
do not intercept at a single density.

In Fig.\,2, the triangles depict for each curve
the temperature $T^* = 0.007 E_F/k_{B}$ for the corresponding density.
In the region confined between  $T=0.05E_F/k_{B}$ and $0.007E_F/k_{B}$,
the exponential dependence  seems to
``saturate'',
but in fact, it crosses over, below
$\approx T^* = 0.007 E_F/k_{B}$, to a  weaker
dependence.
The "high temperature behavior" (for $T > T^*$) of the conductivity
remains metallic-like for all curves in Fig.~2, up to
the highest density studied.
However, the curves taken for high densities
(close to the maximum conductance),
at low temperatures clearly show the onset
of a localizing $\ln T$ dependence
with $ dG/dT >0$. The localizing low temperature dependence
is shown in an expanded scale in Fig.~3a. 
As the temperature is varied, 
it persists for one order of magnitude, 
and does not saturate at low temperatures.  
Its slope $dG/d\ln T \approx 0.35$ is consistent with the
 conventional theory of the weak localization
\cite{altshuler}. Since the ``low-temperature''
localizing $T$-dependence develops on the background of the strong
exponential increase in conductivity at ``high temperatures'',
we conclude, that {\em the exponential raise
in $G (T)$  can not be considered as a proof of the
metallic conductance}, at least
for high densities $n \gg n_{c}$.

The change of sign
of $dG/dT$ shown in Fig.~3\,a for $T<3$\,K
(for $n \approx n_{\rm max}$)
is not caused by significant
changes in disorder for densities around
$n_{\rm max}$. The conductance $G$ (which is
$2\times k_F l$ in the Drude approximation
for the two valley system)
is of the order of  100  ($k_F$ is the Fermi
wave vector and $l$ is the mean free path).
Also, for the spin-orbit parameter
in the chiral model
\cite{lyanda},~  $2 \Delta \tau /\hbar \approx 4 - 8$~ holds
($\Delta$ is the zero magnetic field ``spin-splitting'' at $E=E_F$).
Therefore, the above parameters seem to be not important
at $n \sim n_{\rm max}$.

The picture is less clear for lower densities,
$n \sim (1- 15) \times 10^{11}$ cm$^{-2}$
(see Figs. 3\,b and c),
where the slope decreases, disappear and
finally changes sign to
the negative ``delocalizing''
one $dG/dT <0$ \cite{JETPL98b}.
If the above scenario  would
persist to much lower temperatures,
the conductivity $G(T)$ data taken
for different densities would cross each other
at finite temperatures (but at much lower temperature than 
shown here). This possibility
seems to be unphysical and means that
at least a part of the data taken for lowest temperatures
(most probable, the lower density ones) do not
correspond to the ground state conductivity.
One can not exclude,
therefore, that the low-temperature data
may be affected  by the
tail of the strong metallic-like exponential ``high temperature''
dependence, extending down to low temperatures.

Anyhow, on the basis of the data shown,
it seems rather unlikely, that the
conductivity will grow to infinity
in the $T \rightarrow 0$ limit,
both for high as well as for low carrier densities.
In order to reach a more definite conclusion, the
measurements have to be taken down to
temperatures $T \ll T^* \sim 0.007 E_F/k_{B}$.

In summary, we have found that
the conductivity value in (100) Si-MOS structures
is limited to a finite value, $G_{max} \sim 140$
as density or temperature vary.
We found that the strong metallic-like increase in the
conductivity as $T$ decreases (visible at "high temperatures"
$T > 0.01E_F/k_{B}$)
and the ``low temperature''  behavior (for $T < 0.01E_F/k_{B}$ )
are rather independent of each other.
Despite the observation that the maximum conductivity value
is nearly the same for different Si-MOS samples,
we do not have evidence that this value is related to
a many body ground state \cite{theory}.
The fact that the maximum in $G(T)$ appears at a finite 
temperature ($\sim T^* =0.007E_F/k_{B}$)
indicates actually a single particle origin. 
Such a maximum of $G$ could be the result of a superposition 
of a scattering mechanism and weak localization effects. 
The behavior of the conductivity for
lower temperatures requires further studies.

V.P. acknowledges discussions with B.\ Altshuler, M.\ Baranov,
A.\ Finkel'stein, V.\ Kravtsov, S.\ V.\ Kravchenko,
D.\ Maslov,  A.\ Mirlin, and I.\ Suslov.
The work was supported by RFBR 97-02-17378, by the Programs ``Physics of
solid-state nanostructures'' and ``Statistical physics'', by INTAS,  NWO,
and by FWF P13439, Austria.

  \end{multicols}

\begin{table}
\caption{The parameters of the studied samples. Density is in unites of
$10^{11}$\,cm$^{-2}$.}
\begin{tabular}{|c|c|c|c|c|c|}
sample &
$\mu_{\rm peak} (\mbox{m}^2/\mbox{Vs})$ &
$n_{\rm c,1}$ &
$G_{\rm c,1}$ &
$n_{\rm max}$ &
$G_{\rm max}$ \\
\hline
Si-22 & 3.3 & 0.83 & 0.5 & 39.3 & 140  \\
Si-15a  & 3.2 & 0.82 & 0.4  & 32.1 & 133.7 \\
Si-43b  & 1.96 & 1.4 & 1.5 & 35 & 124.5 \\
Si-4/32 & 0.9 &  2.0 & 1.72 & 61 & 101.2 \\
\end{tabular}
\end{table}

\begin{figure}[tbp]
\caption{ Density dependence of the conductivity for the sample Si22
at 17 different temperatures, $T=0.29$, 1.5, 3.9, 4.8, 5.5, 7.9, 8.5, 10.5,
12.5, 16, 19, 21.5, 23.5, 26, 32, 36, 41\,K. The upper arrow show the density,
$n_{max}$, corresponding to the maximum conductance, the lower arrow is for the
critical density, $n_{c}$.}
 \end{figure}

\begin{figure}[tbp]
\caption{Temperature dependence of the conductivity for  Si-15a
in the range 0.29 to 45K  at 12 density values:
{\it 1}-8.10, {\it 2}- 10.3,
{\it 3}- 15.7, {\it 4}- 21.2, {\it 5}- 26.6,
{\it 6}- 32.1, {\it 7}- 42.94,
{\it 8}- 48.4, {\it 9}- 53.8, {\it 10}- 64.7,
{\it 11}- 75.6, {\it 12}-
$86.5\times 10^{11}$\,cm$^{-2}$.
Continuous lines are for the densities $n<n_{max}$,
dotted lines  for $n>n_{max}$.
The empty triangles depict $T^*$ for the dotted curves {\em 1 - 5},
full triangles are for the continous curves {\em 6 - 12}.}
\end{figure}

\begin{figure}
\caption{Expanded low-temperature part of the
conductivity for the sample Si-43b
in the range 0.29 to 10K for three different density
 values indicated on each panel.  Arrows mark
the temperature $T^* =0.007 E_F/k_{B}$.}
\end{figure}

\end{document}